\documentstyle[aps,twocolumn,floats,epsf]{revtex}

\begin{document}
\wideabs{

\title{A class of exactly solvable pairing models}
\author{J. Dukelsky$^{1}$, C. Esebbag$^{2}$ and P. Schuck$^{3}$}
\address{$^{\left( 1\right) }$Instituto de Estructura de la Materia, CSIC, Serrano 123, 28006 Madrid, Spain.\\
$^{\left( 2\right) }$Departamento de Matem\'aticas, Universidad de Alcal\'a, 28871
Alcal\'a de Henares, Spain.\\
$^{\left( 3\right) }$Institut de Physique Nucl\'eaire, Universit\'e de Paris-Sud, F-91406 Orsay Cedex, France.}

\maketitle

\begin{abstract}
We present three classes of exactly solvable models for fermion and boson
systems, based on the pairing interaction. These models are solvable in any
dimension. As an example we show the first results for fermion interacting
with repulsive pairing forces in a two dimensional square lattice. Inspite
of the repulsive pairing force the exact results show attractive pair
correlations.


\end{abstract}}


Exactly solvable models have played an important role in understanding the
physics of the quantum many body problem, especially in cases where the system is strongly correlated. Such situations arises {\it{e.g.}} in one dimensional (1D) systems of
interest for condensed matter physics and also in strongly correlated finite
fermion systems as atomic nuclei. In both branches of physics the study of
exactly solvable models has been pursued since long with enormous success.

In 1D quantum physics, the exactly solvable models can be classified into
three families. The first family begun with Bethe's exact solution of the
Heisenberg model. Since then a wide variety 1D models has been solved using the Bethe ansatz 
(for a recent review see 
\cite{Ha}). A second family of models are the so called Tomonaga-Luttinger
models\cite{Ha} which are solved by bosonization techniques and which
revealed non Fermi liquid properties in 1D. These systems are now called
Luttinger liquids. The third family, proposed by Calogero and Sutherland,
are models with long range interactions. They have been applied to several
problems\cite{Ha} like spin systems, the quantum Hall effect, random matrix
theory, etc... 

Several exactly solvable models have been developed in the
field of nuclear physics from a different perspective\cite{talmi}. In these
models the hamiltonian is written as a linear combination of the Casimir
operators of a group decomposition chain ideally representing the properties
of a particular nuclear phase. Typical examples are the Elliot SU(3) model
describing nuclear deformations and rotations and the U(6) Interacting Boson Model\cite{iac} with
its three dynamical symmetry limits describing rotational nuclei (SU(3)),
vibrational nuclei (U(5)) and gamma unstable nuclei (O(6)). These
models were extremely useful in providing a simple understanding of some
prototypical nuclei.

The impact of the exactly solvable models in condensed matter physics and in
nuclear physics is so enormous that one hardly can believe that the exact
solution of the Pairing Model (PM), of great interest for both fields,
passed almost unnoticed till very recently\cite{sie1}. The PM, was solved exactly by
Richardson in a series of papers in the sixties\cite{richa0,richa1,richa2}.

Independently of Richardson's exact solution, it was recently demonstrated\cite
{Cambi} that the PM is an integrable model. The pairing model may turn out
particularly interesting, since recent work\cite{Shas} has shown that the
pure repulsive pairing Hamiltonian in a 2D lattice can be solved exactly in
the thermodynamic limit revealing strong superconducting fluctuations. The
importance of this finding stems, of course, from the fact that high $T_{c}$
superconductors apparently acquire their superconducting properties through
the repulsive Coulomb interaction.

We will derive in this letter three families of exactly solvable models
based on the pairing interaction for fermion systems as well as for boson
systems. The most important feature of the new set models we will present is
that they are exactly solvable in any dimension. In\cite{prl} we have
advanced a numerical solution for a three dimensional confined boson
systems, here we will give preliminary results for a fermion system in a 2D
lattice.

Since the proof of integrability and the derivation of the exact solutions
is completely analogous for fermions and bosons, we will develope both
systems in parallel. In what follows whenever there are different signs, the
upper one will correspond to bosons while the lower one to fermions, and we
will refer indistinctly to bosons and fermions as particles.

Let us begin our derivation by defining the three operators

\begin{equation}
\widehat{n}_{l}=\sum_{m}a_{lm}^{\dagger }a_{lm}\quad ,\quad A_{l}^{\dagger
}=\sum_{m}a_{lm}^{\dagger }a_{l\overline{m}}^{\dagger }=\left( A_{l}\right)
^{\dagger }  \label{ope}
\end{equation}
which close the commutator algebra

\begin{equation}
\left[ \widehat{n}_{l},A_{l^{\prime }}^{\dagger }\right] =2\delta
_{ll^{\prime }}A_{l}^{\dagger }~,~\left[ A_{l},A_{l^{\prime }}^{\dagger
}\right] =2\delta _{ll^{\prime }}\left( \Omega _{l}\pm 2\widehat{n}%
_{l}\right)  \label{com1}
\end{equation}

In eq. (\ref{ope}) the pair operator $A_{l}^{\dagger }$ creates a pair of
particles in time reversal states with $%
a^{\dagger }(a)$ the particle creation (annihilation) operator and $\Omega _{l}$
being the degeneracy of level $l$. 

The number operator $\widehat{n}_{l}$ and the pair operators $A_{l}$, $%
A_{l}^{\dagger }$ in each level $l$ close the commutator algebra of the
groups $SU(2)$ for fermions or $SU(1,1)$ for boson systems

The three generators of these algebras can be written in terms of the
previously defined pair and number operators as $K_{l}^{0}=\frac{1}{2}%
\widehat{n}_{l}\pm \frac{1}{4}\Omega _{l}$, and $K_{l}^{+}=\frac{1}{2}%
A_{l}^{\dagger }=\left( K_{l}^{-}\right) ^{\dagger }$.

These generators obey the more familiar commutation relations of the $SU(1,1)$
and $SU(2)$ group algebras 
\begin{equation}
\left[ K_{l}^{0},K_{l^{\prime }}^{+}\right] =\delta _{ll^{\prime
}}K_{l}^{+}~,\quad \left[ K_{l}^{+},K_{l^{\prime }}^{-}\right] =\mp 2\delta _{ll^{\prime
}}K_{l}^{0}  \label{SU}
\end{equation}
The difference between the algebra of $SU(1,1)$ and $SU(2)$ appears in the
sign of the second commutator.

The Hilbert space of $N$ particles moving in $L$ single particle levels can
be classified according to the product of groups $SU\left( 2\right)
_{1}\times SU\left( 2\right) _{2}\times \cdots \times SU\left( 2\right) _{L}$
for fermions or $SU\left( 1,1\right) _{1}\times SU\left( 1,1\right)
_{2}\times \cdots \times SU\left( 1,1\right) _{L}$ for bosons.

A complete set of states in this Hilbert, which are  space spanned by the pair operators (\ref
{ope}), is given by

\begin{equation}
\left| n_{1},n_{2},\cdots ,n_{L},\nu \right\rangle =\frac{1}{\sqrt{{\cal N}}}%
A_{1}^{\dagger n_{1}}A_{2}^{\dagger n_{2}}\cdots A_{L}^{\dagger n_{L}}\left|
\nu \right\rangle  \label{state}
\end{equation}
where ${\cal N}$ is a normalization constant. The possible number of pairs
in each level is $0\leq n_{l}\leq \Omega _{l}/2$ for fermion systems or $%
0\leq n_{l}\leq N/2$ for boson systems. A state $\left| \nu \right\rangle \equiv 
\left| \nu _{1}\nu _{2}\cdots \nu _{L}\right\rangle $
of unpaired particles is defined as

\begin{equation}
A_{l}\left| \nu \right\rangle =0~,~\widehat{n}_{l}\left| \nu \right\rangle
=\nu _{l}\left| \nu \right\rangle ~,~K_{l}^{0}\left| \nu \right\rangle
=d_{l}\left| \nu \right\rangle  \label{senio}
\end{equation}
where $d_{l}=\left( \frac{\nu _{l}}{2}\pm \frac{\Omega _{l}}{4}\right) $ and 
$N=2M+\nu $ , $M$ being the number of pairs and $\nu $ the total number of
unpaired particles. We will borrow from Nuclear Physics\cite{talmi} the name
Seniority for the number of unpaired particles in each level $\nu _{l}.$

In the product spaces mentioned above a model is integrable if there are $L$
independent global operators commuting with one another. These operators are
the quantum invariants and their eigenvalues are the constants of motion of
the system. In looking for these operators, let us define the most general
combination of one and two body operators in terms of the $K$ generators with the
condition of being hermitian and number conserving:

\begin{eqnarray}
R_{l} &=&K_{l}^{0}+\left\{ 2g\sum_{l^{\prime }\left( \neq l\right) }\frac{%
X_{ll^{\prime }}}{2}\left( K_{l}^{+}K_{l^{\prime
}}^{-}+K_{l}^{-}K_{l^{\prime }}^{+}\right) \right.  \nonumber \\
&&\left. \mp Y_{ll^{\prime }}K_{l}^{0}K_{l^{\prime }}^{0}\right\}
\label{Oper}
\end{eqnarray}

Up to now the matrices $X$ and $Y$ are completely free, but we will fix
them imposing the condition that the $R$ operators should commute among one
another to define an integrable model. The condition $\left[
R_{l},R_{l^{\prime }}\right] =0$ will be fulfilled if they are antisymmetric
and satisfy the following equation

\begin{equation}
Y_{ij}X_{jk}+Y_{ki}X_{jk}+X_{ki}X_{ij}=0  \label{cond}
\end{equation}

An analogous condition has been encountered by Gaudin\cite{gaudin} in a spin
model known at present as the Gaudin magnet. His model is based on $R$
operators similar to (\ref{Oper}) but without the one body term. He found
three different solutions for the condition (\ref{cond}) which can be grouped together in compact form as $X_{ll^{\prime }}=\frac{\gamma }{\sin \left[ \gamma \left( \eta _{l}-\eta
_{l^{\prime }}\right) \right] }$, $Y_{ll^{\prime }}= \gamma \cot \left[
\gamma \left( \eta _{l}-n_{l^{\prime }}\right) \right]$, where the different models are distinguished by the value of $\gamma$:

{\bf I. The rational model:} $\gamma \rightarrow 0$

\begin{equation}
X_{ll^{\prime }}=Y_{ll^{\prime }}=\frac{1}{\eta _{l}-\eta _{l^{\prime }}}
\label{A}
\end{equation}

{\bf II. The trigonometric model:} $\gamma=1$

\begin{equation}
X_{ll^{\prime }}=\frac{1}{\sin \left( \eta _{l}-\eta _{l^{\prime }}\right) }%
\ ,\quad Y_{ll^{\prime }}=\cot \left( \eta _{l}-\eta _{l^{\prime }}\right)
\label{B}
\end{equation}

{\bf III. The hyperbolic model:} $\gamma=i$ 

\begin{equation}
X_{ll^{\prime }}=\frac{1}{\sinh \left( \eta _{l}-\eta _{l^{\prime }}\right) }%
\ ,\quad Y_{ll^{\prime }}=\coth \left( \eta _{l}-\eta _{l^{\prime }}\right)
\label{C}
\end{equation}

$\eta _{l}$ is an arbitrary set of non-equal real numbers. Each
solution gives rise to an integrable model and any combination of the $R$
operators produces an integrable hamiltonian. 
Since the three models have quite different properties, we prefer to continue our derivation in an independent way. 
It is worthwile to mention here that if we relax the condition of number conservation on the set of operators (\ref{Oper}) there is a general solution in terms of elliptic functions\cite{gaudin}.      

The rational model has been recently proposed in ref. \cite{Cambi} to
demonstrate the integrability of the PM hamiltonian. Indeed the PM
hamiltonian can be obtained from the rational model by means of linear
combination of $R$ operators $H_{P}=2\sum_{l}\eta _{l}R_{l}^{I}$ plus an
appropriate constant to give

\begin{equation}
H_{P}=\sum_{l}\varepsilon _{l}\widehat{n}_{l}+\frac{g}{2}\sum_{ll^{\prime
}}A_{l}^{\dagger }A_{l^{\prime }}  \label{HP1}
\end{equation}
where the free parameters $\eta _{l}$ have been replaced by the single particle
energies $\varepsilon _{l}\,.$

Then, the PM hamiltonian is diagonal in the basis of common eigenstates of
the $R$ operators. But, also, any function of the $R$ operators defines a
valid integrable hamiltonian. If we want to restrict ourselves to hamiltonians containing at most one and two body terms, the most general linear combination of $R$ operators
is

\begin{eqnarray}
H &=&2\sum_{i}\varepsilon _{i}R_{i}=2\sum_{l}\varepsilon
_{l}K_{l}^{0}+2g\sum_{l\neq l^{\prime }}\left( \varepsilon _{l}-\varepsilon
_{l^{\prime }}\right) X_{ll^{\prime }}K_{l}^{+}K_{l^{\prime }}^{-}  \nonumber
\\
&&\mp 2g\sum_{l\neq l^{\prime }}\left( \varepsilon _{l}-\varepsilon
_{l^{\prime }}\right) Y_{ll^{\prime }}K_{l}^{0}K_{l^{\prime }}^{0}
\label{hgen}
\end{eqnarray}

In (\ref{hgen}) the $X$ and $Y$ matrices can be given formally by one of the
three models (\ref{A}-\ref{C}), and for each of the models they are
functions of an arbitrary set of $\eta ^{\prime }s$ and a pairing strength $%
g.$ Moreover the set of $\varepsilon ^{\prime }s$ in the linear combination (%
\ref{hgen}) is also arbitrary, so that the total number of free parameters
in defining an integrable hamiltonian in each of the three models is $2L+1$
for fermion as well as for boson systems.

We would like to emphasize that the demonstration of integrability given
above for the three models does not imply that they are exactly solvable, $%
i.e.$ that the complete set of common eigenstates of the $R$ operators can
be obtained. As mentioned before, Richardson only worked out the eigenstates
for the PM\ hamiltonian and, as a matter of fact, the authors of ref. \cite
{Cambi} were not aware of the existence of that exact solution. Very
recently \cite{sierra} the eigenvalues of the $R$ operators of the rational
model for fermion systems were obtained, within a conformal field theory
formalism, in the fully paired subspace ($\nu _{l}=0$) and with constant
degeneracy $\Omega _{l}=2$ .

To begin our derivation we will propose an ansatz for the eigenstates of the 
$R$ operators in the Hilbert space of states (\ref{state}) which is a
generalization of the ansatz used by Richardson to find the eigenstates of
the PM. The exact eigenstates for the three models can be written as product
pair wavefunction acting on the space of unpaired particles $\left| \nu
\right\rangle $

\begin{equation}
\left| \Psi \right\rangle =\prod_{\alpha =1}^{M}B_{\alpha }^{\dagger }\left|
\nu \right\rangle , \quad  B_{\alpha }^{\dagger }=\sum_{l}u_{l}\left( e_{\alpha }\right) ~K_{l}^{+} \label{ansa}
\end{equation}

The function $u$ depending on the pair energies $e$, has the form of the
eigenstate of the one pair problem in each of the three models and the set
of parameters $e$, are left as free parameters to be fixed in order to
fulfil the $L$ eigenvalue equations

\begin{equation}
R_{i}\left| \Psi \right\rangle =r_{i}\left| \Psi \right\rangle  \label{eig}
\end{equation}

The collective amplitudes in the pair operators $B\,$for each model are

\begin{equation}
u_{\alpha i}^{I}=u_{i}\left( e_{\alpha }\right) =\frac{1}{2\eta
_{i}-e_{\alpha }}  \label{uA}
\end{equation}

\begin{equation}
u_{\alpha i}^{II-III}=u_{i}\left( e _{\alpha }\right) =\frac{1}{sn\left(
e_{\alpha }-\eta _{i}\right) }  \label{uB}
\end{equation}

We will treat on equal footing the trigonometric and the hyperbolic models
to solve the eigenvalue equation (\ref{eig}). To embody both derivations in
the same formalism we will use the symbols $sn\,$~for $sin$ or $sinh$, $cs$
for $cos$ or $cosh$ and $ct$ for $cot$ or $coth$, not to be confused with
elliptic functions. Note that we have already used $sn$ in (\ref{uB}).

\begin{figure}
\hspace{0.5cm}
\epsfysize=9cm
\epsfxsize=7cm
\epsffile{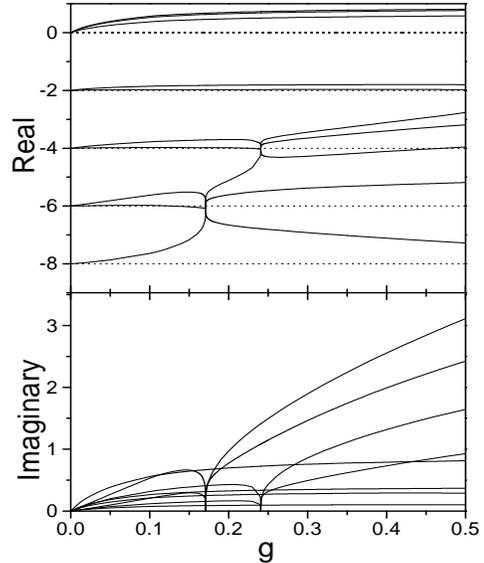}
\narrowtext

\caption{Real and positive imaginary parts of the pair energies $e_\alpha$ for a $6$x$6$ lattice at half filling as a function of $g$. }
\label{fig1}
\end{figure}

Here we summarize the final results for the three models leaving the details
of the derivation for a future publication.

{\bf Model I}

\begin{equation}
1\pm 4g\sum_{j}\frac{d_{j}}{2\eta _{j}-e_{\alpha }}\mp 4g\sum_{\beta \left(
\neq \alpha \right) }\frac{1}{e_{\alpha }-e_{\beta }}=0  \label{richA}
\end{equation}

\begin{equation}
r_{i}=d_{i}\left[ 1\mp 2g\sum_{j\left( \neq i\right) }\frac{d_{j}}{\eta
_{i}-\eta _{j}}\mp 4g\sum_{\alpha }\frac{1}{2\eta _{i}-e_{\alpha }}\right]
\label{eigenA}
\end{equation}

{\bf Models II and III}

\begin{equation}
1\mp 2g\sum_{j}d_{j}~ct(e_{\alpha }-\eta _{j})\pm 2g\sum_{\beta \left( \neq
\alpha \right) }ct(e_{\beta }-e_{\alpha })=0  \label{richB}
\end{equation}

\begin{equation}
r_{i}=d_{i}\left\{ 1\mp 2g\left[ \sum_{j\left( \neq i\right) }d_{j}~ct\left(
\eta _{i}-\eta _{j}\right) -\sum_{\alpha }ct\left( e_{\alpha }-\eta
_{i}\right) \right] \right\}  \label{eigenB}
\end{equation}

In order to obtain the pair energies $e_{\alpha}$, given a set of parameters $\eta$ and a pairing strength $g$, one has to
solve the coupled set of $M$ nonlinear equations (\ref{richA}) for the
rational model or (\ref{richB}) for the trigonometric or hyperbolic models 
respectively. In the limit $g \rightarrow 0$ one imediately realises that 
the equations (\ref{richA}, \ref{richB}) can only be fulfileed for $ e_{\alpha}
\rightarrow 2\eta_i $. Then the amplitudes $u_{\alpha i}$ in (\ref{uA}, \ref{uB})
become diagonal and we see that the states (\ref{ansa}) are equivalent in this limit to the complete set of uncorrelated states (\ref{state}). From this $g \rightarrow 0$ limit one can construct the ground state (with pairs filling the lowest possible states), and the configurations of the succesive excited states. For example, the first excited state is obtained by promoting the highest energy pair to the next upper $2 \eta$ value, or by breaking a pair (removing a pair energy) into two unpaired particles (increasing the senioriy $\nu$ by two). One then follows the trajectory of  the pair energies $ e_{\alpha}$ for each of the configurations as a function of $g$ solving the equations (\ref{richA}, \ref{richB}). 

For bosons systems the pair energies stay always real, but for fermion systems the pair energies can be either real or complex conjugate pairs. In the latter case there might arise singularities in the solution of the equations (\ref{richA}, \ref{richB}) for some critical value of the pairing strength $g_c$ for which two or more pair energies acquire the same value. It can be shown\cite{richmat} that each one of these critical $g$ values is related to a single particle level $i$ and that at the critical point $1-2 d_i$ pair energies must be equal to $2 \eta_i$. These sigularities cancel out in the calculations of energies which do not show any discontimuity in the vicinity of the critical points.                   

\begin{table}
\caption{Single particle energies and degeneracies for the 6x6 lattice.}
\begin{tabular}{drrrrrrrrr}
$\varepsilon _{j}$ & $-4$ & $-3$ & $-2$ & $-1$ & $~0$ & $~1$ & $~2$ & $~3
$ & $~4$ \\   
\hline
$\Omega _{j}$ &  $~2$ &  $~8$ &  $~8$ &  $~8$ & $20$ &  $~8$ &  $~8$ &  $~8$ &  
$~2$ \\ 
\end{tabular}
\end{table}

The eigenvalues of the $R$ operators, given by (\ref{eigenA}) or (\ref
{eigenB}) respectively, are always real since the pair energies are real or complex conjugate pairs. Each solution of the nonlinear set of equations
produces an eigenstate common to all $R_i$ operators, and consequently to any
hamiltonian written as linear combination of them. The corresponding eigenvalue is the linear combination of the $r_i$ eigenvalues. 
As mentioned before, the
most important feature of the three families of models is that they are
exactly solvable in any dimension. The dimensionality enters through the
degeneracies $d_{i}=\nu _{i}/2 - \Omega _{i}/4$. Also through the same
term enters the information about states with nonzero seniority (broken
pairs). Symmetry breaking terms, like anisotropic hoppings or disorder, might lift up the degeneracies but still leaving the hamiltonian as exactly solvable. In such cases the dimensionality would show up in the density of states. 

Next we will present the first results for the rational model of fermions in
a 2D square lattice of $P \times P$ sites with periodic boundary conditions and a repulsive pairing interaction. Assuming a
restricted hopping term between nearest neighbors, the single fermion levels
are $\varepsilon _{k}=-2\left( \cos k_{x}+\cos k_{y}\right)$, with $%
k_{\sigma }=2\pi n_{\sigma }/P$ and $-P/2\leq n_{\sigma } < P/2$.
 Here we will consider a $6\times 6$ lattice at half filling ($M=18$) with a PM hamiltonian for which $\eta _{i}=\varepsilon
_{i}$ in (\ref{richA}), but the properties we will discuss are quite general
and independent of the latter choice.

The single fermion energies $\varepsilon
_{k}$ and the corresponding degeneracies $\Omega _{k}$ are displayed in the Table I.  In the limit $g=0$ the groundstate is
obtained by distributing the $M=18$ pairs in the lowest possible states.

In Fig. 1 we show the real and imaginary part of the pair energies. Only real part and the positive imaginary part are shown for each complex conjugate pair. The first level accomodates one pair which is forced to be real. The next three levels accomodate four pairs each, forming two pairs of complex conjugates in each level. In the last level we can put five pairs of particles forming two pairs of complex conjugates and one real pair energy.   
For the critical value $g_1 = 0.1708$ the first five pairs become equal to $2 \varepsilon_2=-6$ as discussed before. As seen in Fig. 1, the real part of the two first complex conjugated pairs together with the first real pair energies cross at $-6$ while the imaginary parts go to zero.
A similar situation arises for the second critical point at $g_2=0.2407$ in which five pairs become equal to $2 \varepsilon_3 = -4$. 

\begin{table}
\caption{Correlation energy densities and Interaction energy densities  for various values of $g$ for the 6x6 lattice at half filling.}
\begin{tabular}{drrrrrr}
$g$ & $0.0$ & $0.1$ & $0.2$ & $0.3$ & $0.4$ & $0.5$ \\
\hline
$E _{corr}$ & $0.0$ & $-0.74$ & $-1.55$ &  $-2.37$ &  $-3.19$ &  $-4.03$ \\
\hline
$E _{int}$ & $0.0$ & $0.12$ & $0.18$ &  $0.22$ &  $0.25$ &  $0.28$ \\
\end{tabular}
\end{table}

In spite of the quite involved behavior of the pair energies, the total energy displays a smooth behavior as a function of $g$. 
In Table II we give some values of the correlation energy density ($(E(g)-E_{HF}(g))/P^2$) and the interaction energy density ($(E(g)-E(0))/P^2$) as a function of $g$. The interaction energy increases with increasing values of $g$, showing a tendency to saturation that is consistent with its vanishing in the thermodynamic limit \cite{Shas2}. The correlation energy density decreases almost linearly with $g$.   
These
attractive pairing correlations based on a repulsive interaction give further
numerical support to the work of Shastry\cite{Shas} who found quasi-long ranged order in the thermodynamic limit of the model at half filling.

In summary, we have presented three families of new exactly solvable
models based on the pairing interaction for fermion and boson
systems. These models have the important feature of
being solvable in any dimension. We have presented preliminary results
for the properties of the exact solution of the rational model
in a 2D square lattice with repulsive pairing interaction. This model
may be useful to study features of high Tc superconductivity
because, inspite of the purely repulsive bare force, the exact solution shows
attractive pair correlations.

After completing this work we learned of a recent
preprint\cite{grain} in which the hyperbolic model is presented for electrons in ultrasmall superconducting grains, however without
applications. This model is equivalent to our models II and III for fermions in 1D with
seniority zero.

{\bf Acknowledgments} This work was supported by the DGES Spanish grant
BFM2000-1320-C02-02.


\end{document}